\documentstyle[prl,preprint,aps]{revtex}
\begin{document}
\draft
\title{Composite Fermion Description of Correlated Electrons in Quantum
Dots: Low Zeeman Energy Limit}
\author{R.K. Kamilla and J.K. Jain}
\address{Department of Physics, State University of New York
at Stony Brook,\\ Stony Brook, New York 11794-3800}
\date{\today}
\maketitle
\begin{abstract}
We study the applicability of composite fermion theory to
electrons in two-dimensional parabolically-confined quantum dots
in a strong perpendicular magnetic field in the limit of low Zeeman energy.
The non-interacting composite fermion spectrum correctly specifies the primary
features of this system. Additional
features are relatively small, indicating that the
residual interaction between the composite fermions is weak.
\footnote{Published in Phys. Rev. B {\bf 52}, 2798 (1995).}
\end{abstract}
\section{Introduction}

Recent progress in microlithography \cite {Reed} has made it
possible to fabricate
artificial semiconductor structures containing only a few electrons, called
quantum dots. Electrons can be added to a quantum dot
one by one, and its properties can be investigated by various
techniques, e.g., tunneling \cite {Expt1,Klein}, capacitance measurements
\cite {Ray}, and optical spectroscopy \cite {Heitman}.
In particular, the chemical potential $\mu_{N}$ of the $N$
electron system can be measured as a function of various parameters,
e.g. the confining potential or
the external magnetic field $B$. The chemical potential
is given by $\mu_{N}=E_{N}-E_{N-1}$, where $E_{N}$ is the
ground state energy of the $N$ electron system, and consequently
contains information about the ground state energy of interacting electrons.
As some
parameter (say $B$) is changed, the ground states of the $N$ and $N-1$
electrons vary continuously until a level crossing occurs, at which time
$\mu_{N}$ exhibits a cusp
\cite {Yang,Chaplik,Hawrylak,Palacios,Daniela1,Chamon,Lai};
 such cusps are observed in various experiments \cite
{Expt1,Klein,Ray,Heitman,McEuen}.

A completely non-interacting electron model is not sufficient for
describing the quantum dot physics, and the repulsive Coulomb interaction
must be taken into account at some level. In the simplest scenario,
the energy to add an extra electron can be modeled in terms of a
classical capacitance, which is a smooth function of the number of
electrons in the dot \cite {Averin}. Superimposed over this smooth classical
contribution are small fluctuations, that originate from either
the quantization of single-particle energy levels
in the quantum dot, or correlations due to the Coulomb
interaction, or a combination of the two.
In the zeroth order approximation, the corrections to the
classical energy may be computed in terms of non-interacting
electrons in the quantum dot \cite {McEuen}.
Then, whenever there is a level crossing in the non-interacting
ground state, a cusp appears in $\mu_{N}$. At high
$B$, a level crossing occurs when an electron changes its
Landau level (LL) index, or reverses its spin within
the same LL. Klein {\em et al.} \cite {Klein}
have investigated in detail the region where both spin
species of only the lowest LL are occupied, and found that a Hartree-Fock
theory \cite {Yang,Chamon,Daniela}
provides a reasonably accurate quantitative account of the
cusp positions; the cusps in this
case originate when the electrons flip their spin one by one, until
they are fully polarized.

At still higher $B$, when the filling factor $\nu<1$, and all
electrons are fully polarized, an interplay of different LL's is not
possible.  A non-interacting electron model or
a Hartree-Fock-type calculation will obtain the ground state energy
to be a smooth function of various parameters, and as a result will
not produce any level crossings.  However,
detailed exact diagonalization studies on small systems show that
level crossings do indeed occur in this regime
\cite {Yang,Chaplik,Lai,Girvin,Johnson}. These originate
exclusively from Coulomb correlations.
Phase diagrams \cite {Yang,Chaplik}
for the ground state as a function of the
magnetic field and the strength of the confinement have been
constructed using exact diagonalization studies. Such studies,
however, are possible only for quantum dots with very few electrons.

The results of exact
diagonalization studies of fully polarized electronic states have been
interpreted in terms of the composite fermion theory \cite
{Beenaker,Kawamura}.
The motivation comes from the relevance of composite fermions to the
phenomenon of the fractional quantum Hall effect (FQHE) \cite
{Stormer} that occurs when two-dimensional electron systems (2DES)
are exposed to very high magnetic fields.
In this framework, the FQHE can be understood
as the Integer Quantum Hall Effect
(IQHE) of composite fermions \cite {Jain}.
The FQHE occurs as a result of incompressibility, i.e.,
cusps in the ground state energy as a function of the magnetic field,
which suggests that the physics of the cusps in quantum dots is
related to the physics of the FQHE. It was shown that the
cusps in the quantum dot states can be
understood in terms of non-interacting composite fermions \cite
{Beenaker,Kawamura}. Their origin is briefly as follows. The electron
system completely confined to the lowest LL maps on to composite
fermions with several {\em quasi-}LL's occupied, and the interaction
energy of electron is mapped into an effective cyclotron energy of the
composite fermions. Level crossings in $E_{N}$ then occur as a result
of an interplay between various quasi-LL's of composite fermions,
i.e., as the composite fermions change their quasi-LL index one by one.
Thus, the CF theory effectively provides a single particle description
for the correlation effects, and gives a very
simple intuitive picture for understanding the principal features of
the strongly correlated quantum dot ground states.

The study of Refs. \cite {Beenaker,Kawamura} extended the
applicability of the CF model to systems with non-uniform densities,
and edges. (Most earlier studies used the spherical geometry which has
no edges.) Also, since the composite fermions are many-body
objects, one may ask how many electrons are needed before
the CF description becomes valid. (Of course, there is no composite
fermion for a single electron.) In the case of fully polarized
electrons, the CF description was found to be reasonably good
even for as few as three electrons \cite {Kawamura}.

This work investigates the applicability of the CF theory to quantum
dots in the limit of vanishing Zeeman energy. This limit is
relevant (in certain parameter range) because the Zeeman splitting
is rather small in GaAs, roughly 1/60 of the Cyclotron energy, since
the band g-factor ($g^{*} \sim 0.44$) of electrons and their band
mass ($m^{*} \sim 0.067 m_{e}$) are both very small.
Experimental \cite {Willet} and theoretical \cite {Halperin} studies
in FQHE have shown that even for moderately strong magnetic fields,
the ground-states are not always fully spin-polarized (e.g., $\nu={2
\over 3}, {2 \over 5},{2 \over 7}$, etc. have spin-singlet ground-states).
Similar physics can be expected in
quantum dots. Certain features in single-electron capacitance
spectroscopy \cite {Ray} and tunneling \cite {Expt1,Klein}
experiments in quantum dots have been interpreted in terms of spin
singlet states, and the effect of the spin degree of
freedom has been investigated in several theoretical
studies \cite {Yang,Chaplik}.

In nonfully-polarized states electrons may overlap spatially and
hence one would
expect the residual interaction between composite fermions to be
of greater importance than for fully polarized composite
fermions \cite {XGW}. However, we find that the non-interacting
CF model still identifies the relatively strong cusps.
The additional weaker cusps are a signature of a residual
interaction between composite fermions. While the true residual
interaction between the composite fermions is not known,
even  a minimal ``delta-function'' hard-core interaction gives
reasonable qualitative picture for the weaker structure.
The low-Zeeman-energy quasi-LL gap is found to be
roughly an order of magnitude
smaller than that of the fully polarized composite fermions.
For example, in units of
${e^2 \over \varepsilon a}$, where $a$ is the effective magnetic length,
the quasi-LL gap in the large Zeeman energy limit
\cite {Kawamura} was $\sim 0.12 \;(N=5)$,
whereas in the small Zeeman energy limit
the gap is $\sim 0.019 \;(N=5)$.
Interestingly, we find that the CF description improves as the
number of electrons confined in the dot increases.

Our main objective here is to investigate the validity of
the CF model, for which we will compare its consequences with
the results of exact diagonalization studies. Therefore, we restrict
our study to small systems. It should however be emphasized that the
details of the CF model may be worked out  rather
straightforwardly even for a large number of electrons.
The non-interacting CF model should therefore
prove useful in analyzing experimental
results in the high field regime where the Hartree-Fock approximation
is not applicable, and exact diagonalization cannot be performed.

The plan of this paper is as follows.
Section II gives the results of exact numerical diagonalization.
Section III compares the CF theory with numerical studies at zero
Zeeman energy.  In section IV, we consider the electron-ground-state
phase diagrams as a function of the confinement strength and magnetic
field, and also some typical addition spectra.

\section{Numerical Calculations}

We will consider two-dimensional quantum dots
with parabolic confinement $(1/2)m^{*} \omega_o|{\bf r}|^{2}$
\cite {Fock}.
Further, we will assume sufficiently high $B$ that
the cyclotron energy $\hbar\omega_c$ is large compared to
the confinement energy $\hbar\omega_o$, but the Zeeman
energy remains sufficiently small (which will be set to zero
throughout this work).
The total angular momentum $L$ commutes with the Coulomb interaction,
which makes it possible to diagonalize the problem in various $L$
subspaces separately. For zero Zeeman energy, the total spin
$S$ is also a good quantum number, so it
is sufficient to work in the subspace of the lowest $ S_z $,
provided it is remembered that each
state in this subspace represents a multiplet of $2S+1$
degenerate states of the full Hilbert space.
In the high $B$ limit, the ground state energy $E(L)$ in a given $L$
subspace nicely separates
into two parts, the confinement energy $E_{c}(L)$ and the interaction
energy $V(L)$. The former is given by
\begin{equation}
E_c = {\hbar\omega_c \over 2}\left [ {\left (1+4{\omega_o^2 \over
\omega_c^2} \right )}^{1 \over 2}- 1\right ]L\;\;,
\end{equation}
and $V(L)$ is the same as the interaction energy of the ground state
at $L$ {\em without} any confinement, provided the magnetic length
is replaced by an effective magnetic length $a$, given by
$a={\sqrt{\hbar \over m^{*}}}
{\left ( \omega_c^2 + 4\omega_o^2  \right )}^{-{1 \over 4}}$. As a
result, it is sufficient to calculate $V(L)$ for a system without any
confinement.

The true ground state of the system is determined by the minimum of
$E=E_{c}+V$. While it will depend on various
parameters, like $B$ or $\omega_{o}$, it is in general a state where
the plot of $E(L)$ vs $L$ has a downward cusp; states
with upward cusps will never become ground states. It is convenient to
define the size of a cusp at $L$ as
$\Delta(L) = E(L+1)+E(L-1)-2E(L)$. Then, only states with positive
$\Delta$ may become ground states with suitable choice of parameters.
Since $E_{c}$ is a linear function of $L$, $\Delta =
V(L+1)+V(L-1)-2V(L)$, and the possible ground states can be identified
directly from the plot of $V(L)$.

The Coulomb Hamiltonian was diagonalized exactly for Hilbert spaces of
sizes less than 3500, using standard numerical subroutines.
For bigger Hilbert spaces, we used
a modified Lanczos technique \cite {Gagliano} to obtain a few of the
low energy states. The maximum size we have studied is approximately
45000. The Lanczos scheme requires care in dealing with
states that are almost degenerate.
Exact diagonalization has been carried out in a number of earlier
studies, to which the reader is referred for details \cite
{Yang,Chaplik,Hawrylak,Palacios,Daniela1,Lai,Girvin,Johnson,Stone,Laughlin,Tapash}.

Fig.~(1) shows $V(L)$ as a function of $L$ for 3-6 electrons.
The structure on the curves is small compared to that seen in
the analogous plot for spin-polarized electrons
\cite {Kawamura}. In order to bring out the cusps more clearly,
we plot in Figs.~(2-5) $\Delta(L)=V(L+1)+V(L-1)-2V(L)$ as a
function of $L$. As stated earlier, $\Delta$ is positive
for downward cusps and negative for upward cusps.

\section{Composite Fermion Description}

Composite Fermions are relevant when
two-dimensional electrons are subjected to a strong magnetic field.
The essential role of the Coulomb interaction is presumed
to bind an even number of
vortices of the many-particle wavefunction to each electron. The
resultant electron + vortex combination has the statistics of a
fermion and is called a composite fermion:
\begin{equation}
        electron + 2$m$\;*\;vortices \longrightarrow composite\; fermion\;.
\end{equation}
Microscopically, the formation of composite fermions implies that the
(unnormalized) low-energy wave functions of interacting
electrons with total angular momentum $L$ have the form:
\begin{equation}
\Psi_{L}={\cal P}\prod_{j<k}(z_{j}-z_{k})^{2m}\Phi_{L^*}\;\;,
\label{cfwf}
\end{equation}
\begin{equation}
L=L^*+mN(N-1)=L^*+2mM\;\;,
\label{cfL}
\end{equation}
\begin{equation}
M={N(N-1) \over 2}
\end{equation}
where $\Phi_{L^*}$ is the wave function of
electrons with total angular momentum $L^*$, and
${\cal P}$ projects the wave function on to the LLL of
electrons, as appropriate for $B\rightarrow\infty$.
The Jastrow factor $\prod_{j<k}(z_{j}-z_{k})^{2m}$
binds $2m$ vortices to each electron of
$\Phi_{L^*}$ to convert it into a composite fermion.
Non-interacting composite fermions are obtained when $\Phi_{L^*}$ is
taken to be the  wave function of non-interacting electrons.
The LL's of non-interacting electrons of $\Phi_{L^*}$
map into quasi-LL's of composite
fermions, separated by a quasi-cyclotron energy gap, which is treated
as a parameter of the theory. The system of
interacting electrons at arbitrary $L$
is mapped into a system of composite fermions in the range $-M\leq
L^* \leq M $ with a suitable choice of $m$. $M$ is the angular momentum
of the $\nu=1$ state of fully polarized electrons.
Note that the above discussion is applicable in both the large and
the small Zeeman energy limits. In the former, the electrons in
$\Phi_{L^*}$ are assumed to be fully polarized, whereas in the latter,
which is the case in the present study, electrons in $\Phi_{L^*}$
can have either spin, with the Zeeman energy set to zero.

\subsection{Non-interacting composite fermions}

According to the CF theory, the interacting electron system at $L$ is
equivalent to a weakly interacting CF system at $L^*=L-2mM$.
In Fig.~(6) we also plot the kinetic energy of composite fermions as a
function of $L^*$. This is the same as that of non-interacting
electrons at $L^*$, but with the cyclotron energy of electrons
replaced by an effective cyclotron energy of composite fermions,
denoted by $\hbar\omega^{CF}$, which is to be determined emperically
by comparison with the interaction energy curve.

It is worthwhile to consider in some detail why cusps appear in
Fig.~(6). Consider $N=4$. For non-interacting electrons, there are no
cusps for $L\geq 2$; at $L=2$ the occupied single particle states in
the lowest LL are $0\uparrow$, $0\downarrow$, $1\uparrow$, and
$1\downarrow$, where $j\downarrow$ denotes the single particle state
with angular momentum $j$ and spin down. The exact diagonalization
calculation for interacting electrons however shows cusps. Let us go
to the non-interacting CF basis.
The total angular momentum of composite fermions is $L^*=L-12$.
All composite fermions can be accommodated in the lowest quasi-LL for
$L^*\geq 2$, which corresponds to $L\geq 14$. As $L^*$ is reduced, one
composite fermion must be pushed into the second quasi-LL. The lowest
$L^*$ with only one composite fermion in the second quasi-LL is
$L^*=0$, where the ground state contains three composite fermions
in the lowest quasi-LL, in angular momentum states  0,0, and 1, and one
in the second quasi-LL, in the angular momentum state -1 \cite
{footnote}. For
decreasing $L^*$ further, the CF-kinetic energy must be raised
further.  This would result in a cusp in the interacting electron
system at $L=12$.  Similar analysis produces the curves of Fig.~(6).

The cusp sizes are plotted in
Figs.~(2-5). The prominent cusps in the interaction energy
curve are well reproduced by the non-interacting-CF theory.
The effective cyclotron energy, in units of ${e^2
\over \varepsilon a }$, is determined to be
$\sim 0.012 (N=3),\; 0.017 (N=4),\; 0.019 (N=5)$ and $0.017 (N=6)$.
It is interesting that the description of interacting
electrons in terms of composite fermions becomes better for larger
$N$. For example, the negative CF cusps
for $N=5$ and $N=6$ correspond to  negative cusps in
the interacting energy spectrum, but to positive cusps for
$N=3$ and 4 (even though the overall shape is obtained correctly).
This is not surprising, since composite fermions are inherently
many-body objects, and may
not be appropriate for systems with very few electrons.
For zero Zeeman energy, the composite fermion description is only
qualitatively valid for quantum dots with fewer than five electrons.

\subsection{Residual interaction between composite fermions}

There are additional cusps in the $V(L)$ curve, of relatively small
sizes, that cannot be explained within the non-interacting CF model.
In particular, this model predicts an absence of
cusps in the region corresponding to  $M^{\prime}\leq |L^*| \leq M$,
where $M^{\prime}$ is the smallest angular momentum
possible within the lowest LL, but there are several cusps in $V(L)$
in the corresponding region $ M^{\prime} + 2mM \leq L \leq
(2m+1)M $.  The additional cusps can be interpreted as
originating from a residual interaction between the composite
fermions.  These cusps are roughly an order of magnitude weaker than
the primary cusps, showing that the residual interaction between
the composite fermions is weak compared to their effective cyclotron
energy.

The residual interaction between the composite fermions may be
incorporated by taking the composite fermions at $L^*$ to be
interacting. We have considered two models in an attempt to mimic the
residual interaction, one in which the composite fermions at $L^*$
interact via the Coulomb interaction, and the other in which they
interact via a hard-core interaction.
For simplicity, we have
considered only the range $M^{\prime}\leq L^*\leq M$, where all
composite fermions are in the lowest quasi-LL.
Figs.~(2-5) show the Coulomb interaction energy cusp-size (panel a,
broken lines). It is clear
that it captures the qualitative physics of the weaker cusps.
The same is true of a  hard-core delta function interaction. This
suggests that the dominant part of the residual interaction is the
contact interaction, explaining
why the residual interaction is less important
for fully polarized electrons.

\subsection{Low energy spectrum}
Besides the shape of the $V(L)$ curve, the CF scheme also sheds
light on the low-energy spectrum of states of interacting electrons.
Fig.~(7a) shows the low-energy spectrum of a five electron system in the
range $24 \leq L \leq 30$. The spin quantum number is shown on the figure
for a few of the low-energy states.
Fig.~(7b) shows the spectrum of composite
fermions in the range $4 \leq L^* \leq 10$, interacting
via the Coulomb interaction. The low-energy spectrum of Fig.~(7b) matches
nicely that of Fig.~(7a). At several values of $L$, there are two nearly
degenerate ground states in Fig.~(7a). This is also seen in
Fig.~(7b), although the ordering of the two states is sometimes reversed.
Note that it is crucial to consider {\it interacting} composite fermions
in order to explain the low-energy spectrum - for non-interacting
composite fermions, all states at a given $L^*$ in Fig.~7(b)
would be degenerate.

The ground state at $L=(2m+1)M$ is fully spin-polarized, known to be well
described by the Laughlin wave function \cite {Laughlin}.
 We find that there is
another state very close to it in energy. We do not fully understand the
origin of this state at the moment. It is believed that the Laughlin
state is non-degenerate in the thermodynamic limit even for zero Zeeman
energy, except for the spin multiplicity. In our calculations, we find
that the energy difference between the two states increases with $N$
( it is 0.0027, 0.0046 and 0.0047, in units of ${e^2 \over \varepsilon a}$
for $N=4,5$ and $6$, respectively),
suggesting that the other state may not be relevant in the thermodynamic
limit. We also
note here parenthetically that the ground state quantum numbers in our
calculations are sometimes in disagreement with those of ref. \cite {Yang}.

\subsection{Nature of the ground states}

It was noted in ref. \cite {Kawamura} that the ground states
of interacting electrons are ``compact" states of composite fermions.
In order to define compact states, and their
relationship to ground states, let us
first consider the system of non-interacting fermions in the range
$-M\leq L^* \leq M$.
Let us fix the number of electrons in the $k^{th}$ LL to be $N_{k}$
($\sum{N_{k}} = N$). Then, the state with the
smallest total angular momentum is called compact, denoted by
$[N_{0},N_{1},...  N_{K}]$, where $K$ is the index of the
highest occupied LL. This state has the property that the $N_{k}$ electrons
in the $k$th LL occupy the innermost angular momentum states (hence
the name compact). It is clear that all
 of the positive cusp states of non-interacting
fermions are compact, since otherwise the total angular momentum can
always be decreased without increasing the kinetic energy.
(Of course, every compact state is not necessarily associated with a
positive cusp.)

We have already shown that the principal cusps of interacting
electrons can be explained in terms of non-interacting composite
fermions. These states are therefore compact states of composite
fermions. Their wave functions, denoted by
$[N_0,N_1,\ldots,N_K]_{CF}$, are obtained
according to Eq.\ (\ref{cfwf}), i.e.,  by  multiplying the wave
functions of compact electron states by
the Jastrow factor, and then projecting on to the lowest LL. These wave
functions have been tested successfully for the case of fully
polarized electrons \cite {Kawamura,Dev}.

Note that the $L^{*}$ spectrum
from $0$ to $M$ is related by reflection symmetry to the other half
($0$ to $-M$). The horizontal lines of one side map into the tilted lines
of the other side and vice-versa, and a cusp at $L^{*}$ also implies a
cusp at $-L^{*}$.  Similar approximate mirror symmetry
is also seen here in Figs.~(2-5) about $L = 2mM$ in
the interacting electron ground-states.
In particular, the spin quantum numbers
and the cusp sizes are (approximately) reflected about $L = 2mM$.

	In the disk geometry, all cusps may not be associated with
thermodynamic FQHE states, as indicated in ref. \cite {Kawamura}.
This is related to the fact that the IQHE state at $\nu^{*} = n$
cannot in general be identified precisely. For fully spin-polarized
electrons, the only exception is the $\nu^{*}=1$ state, which allows an
identification of the Laughlin states at $\nu={1 \over (2m+1)}$; the
other cusps are not associated with any thermodynamic FQHE states.
In the present case, with low Zeeman energy, the IQHE state at $\nu^{*}
=2$ is the only state that can be identified unambiguously: for an
even number of electrons, it occurs at $L^{*}={N \over 2}({N \over 2} -1)$.
 Consequently, the cusp at
$L=2M + {N \over 2}({N \over 2} -1)$ can be a
ssociated with the ${2 \over 5}$ spin-singlet FQHE
state, and the cusp at $L=2M-{N \over 2}({N \over 2} -1)$
with the ${2 \over 3}$ spin-singlet
state. These are the last and the first cusps in the CF-kinetic energy
curves. Other cusps do not correspond to FQHE states.

\section{Conclusion}

This study analyzes interacting lowest-LL electrons
in quantum dots in the small Zeeman energy limit
in the framework of composite fermions, and shows that the
prominent features are understood in terms of non-interacting
composite fermions. For
more detailed information, it is necessary to incorporate the
residual interaction between composite fermions. While the true interaction
between composite fermions is necessarily quite complex, we find that any
repulsive two-body interaction provides a reasonable qualitative picture.

We thank T. Kawamura, X.G. Wu, V.J. Goldman, E. Yang, R. Ashoori and O. Klein
for helpful discussions and communications. This work was supported in
part by the Office of Naval Research under Grant no. N00014-93-1-0880.

{\bf Figure Captions}

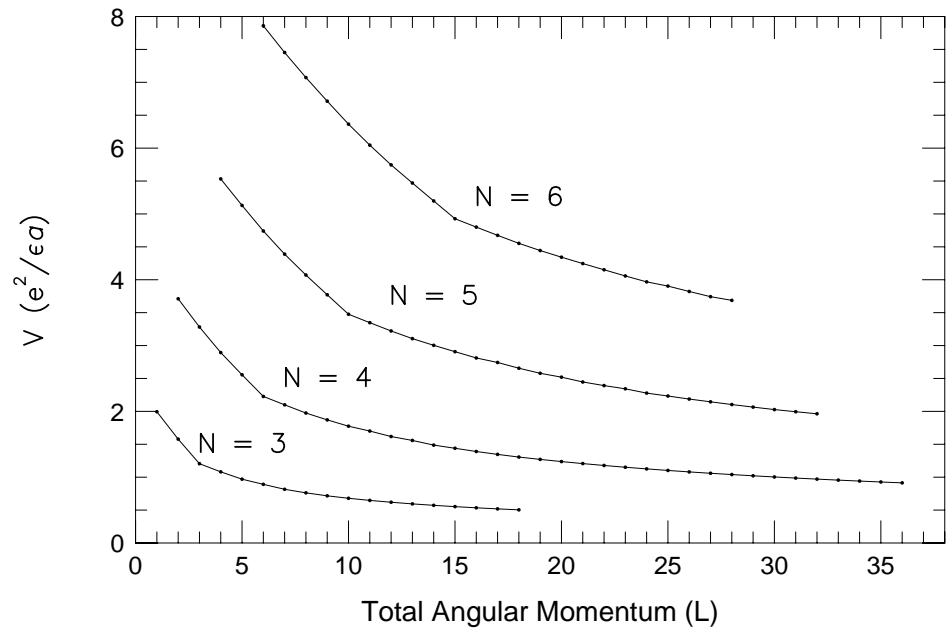
Fig. 1. The interaction energy of the ground state in the $L$
subspace, $V(L)$,
as a function of the total angular momentum $L$. $N$ is the
number of electrons.

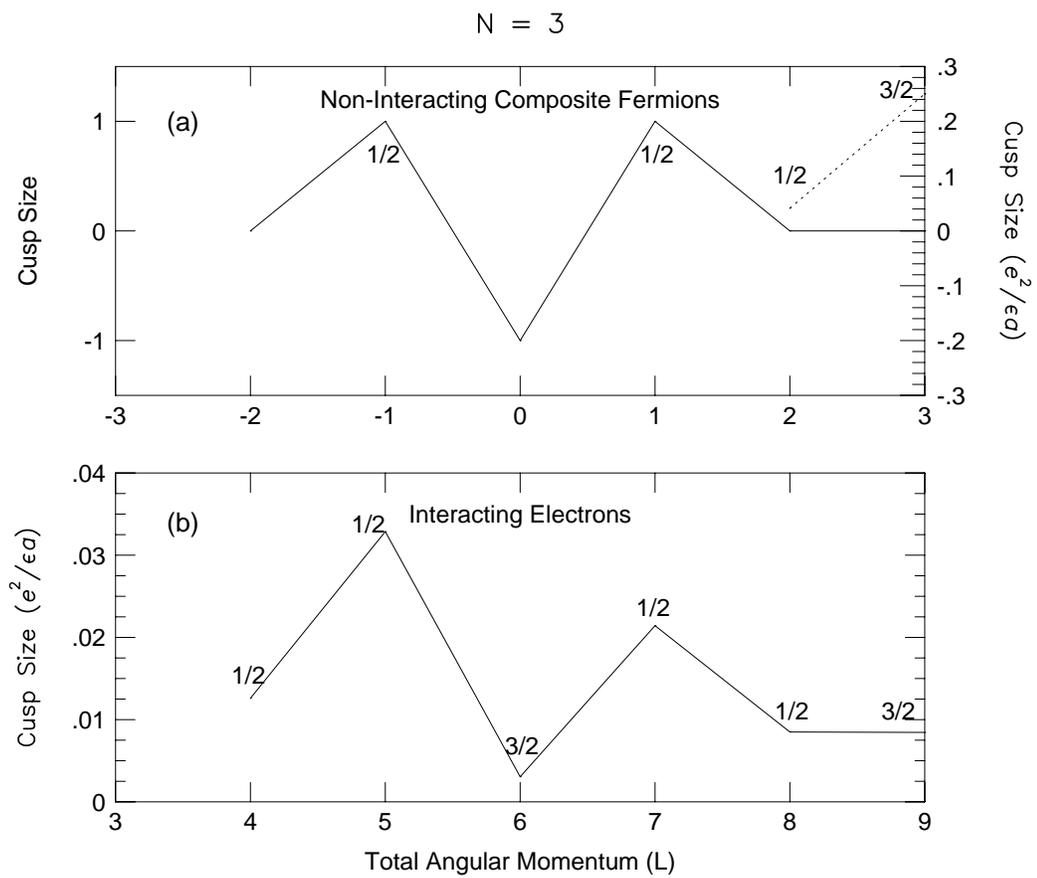
Fig. 2. The cusp size $\Delta(L)$, defined in the text,
for non-interacting
composite fermions (upper panel, solid line, left scale) and
interacting electrons (lower panel, right scale). The dashed line in
the upper panel shows the cusp size for interacting composite
fermions (right scale). The $\Delta$ of non-interacting composite fermions
is given in units of ${\hbar \omega^{CF}}$; the rest are in units
of ${e^2 \over \varepsilon a}$. The ground state spins are
shown on the plot itself. For non-interacting composite fermions,
spins are not shown when there are many degenerate ground states; for
interacting electrons (or interacting composite fermions),
two spins are shown when the first excited state  is almost degenerate
with the ground state (the lower spin corresponds to the actual
ground state).

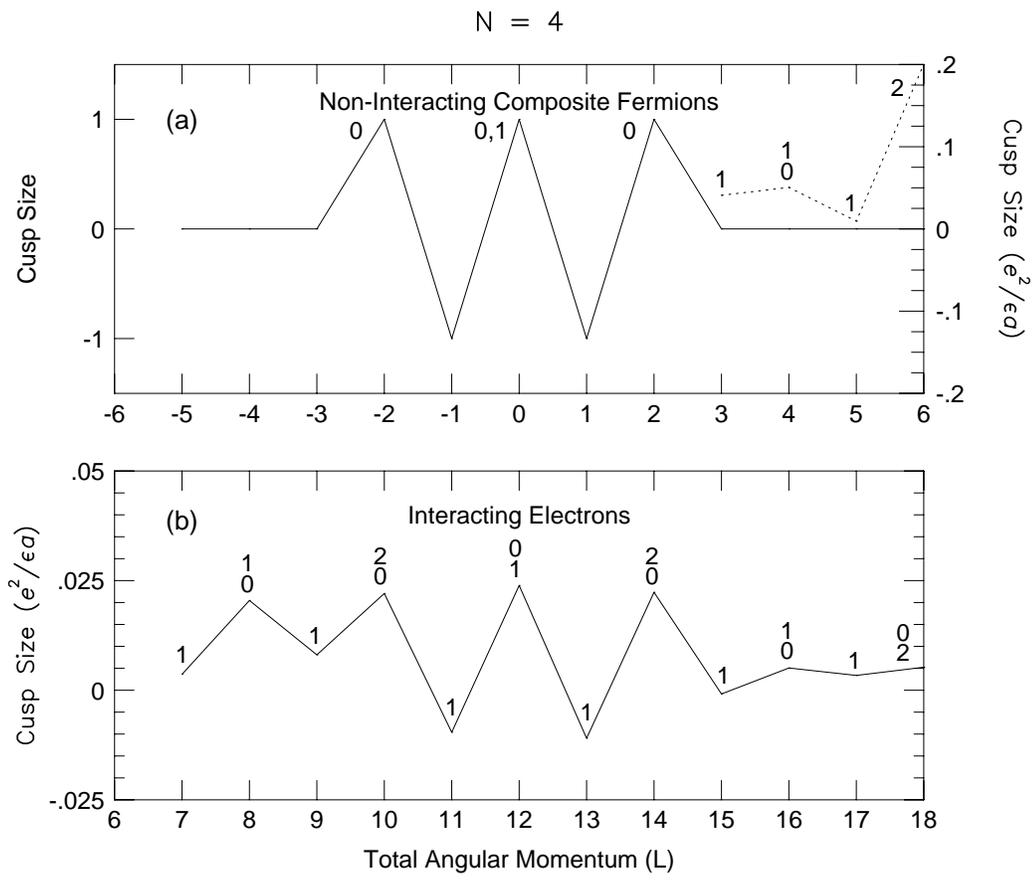
Fig. 3. Same as in Fig. 2 for $N=4$.

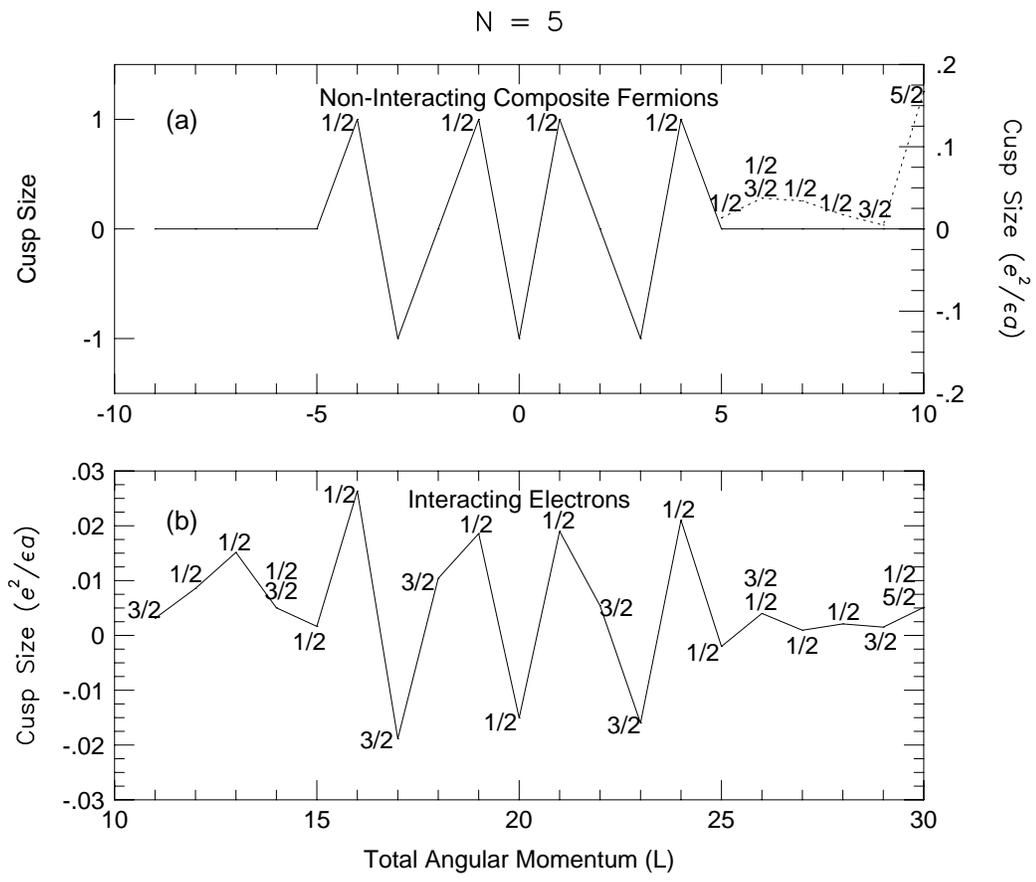
Fig. 4. Same as in Fig. 2 for $N=5$.

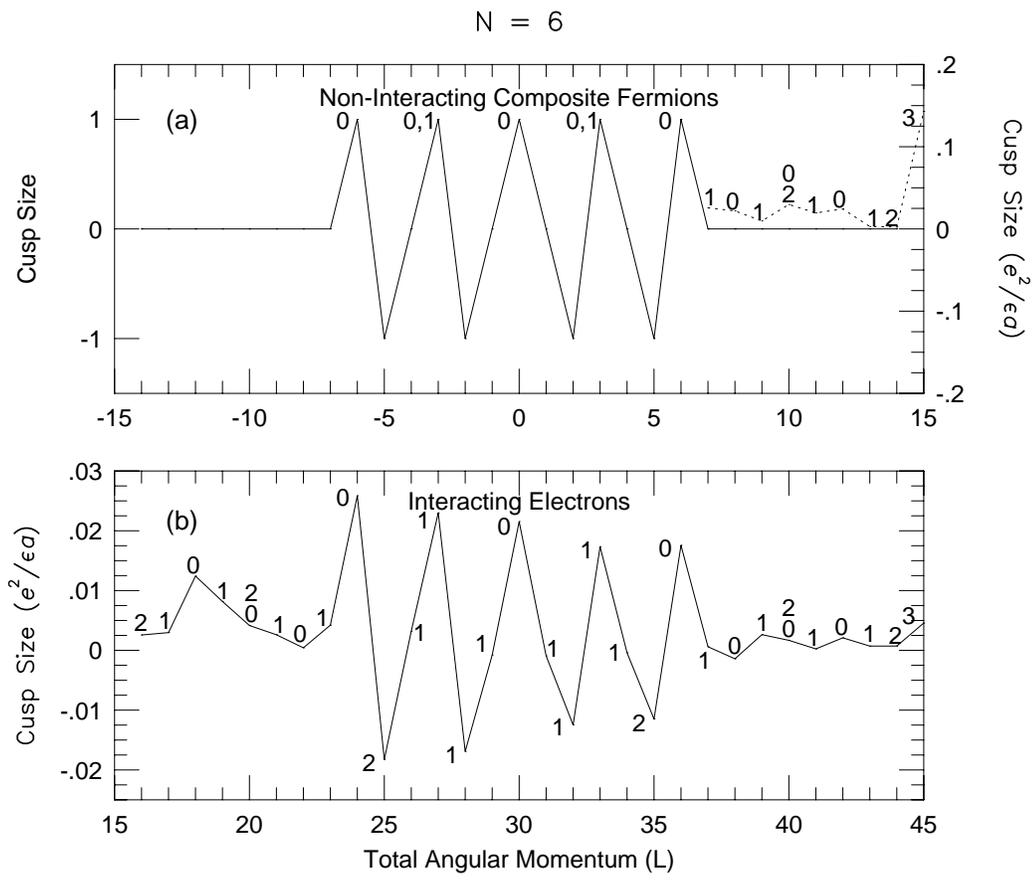
Fig. 5. Same as in Fig. 2 for $N=6$.

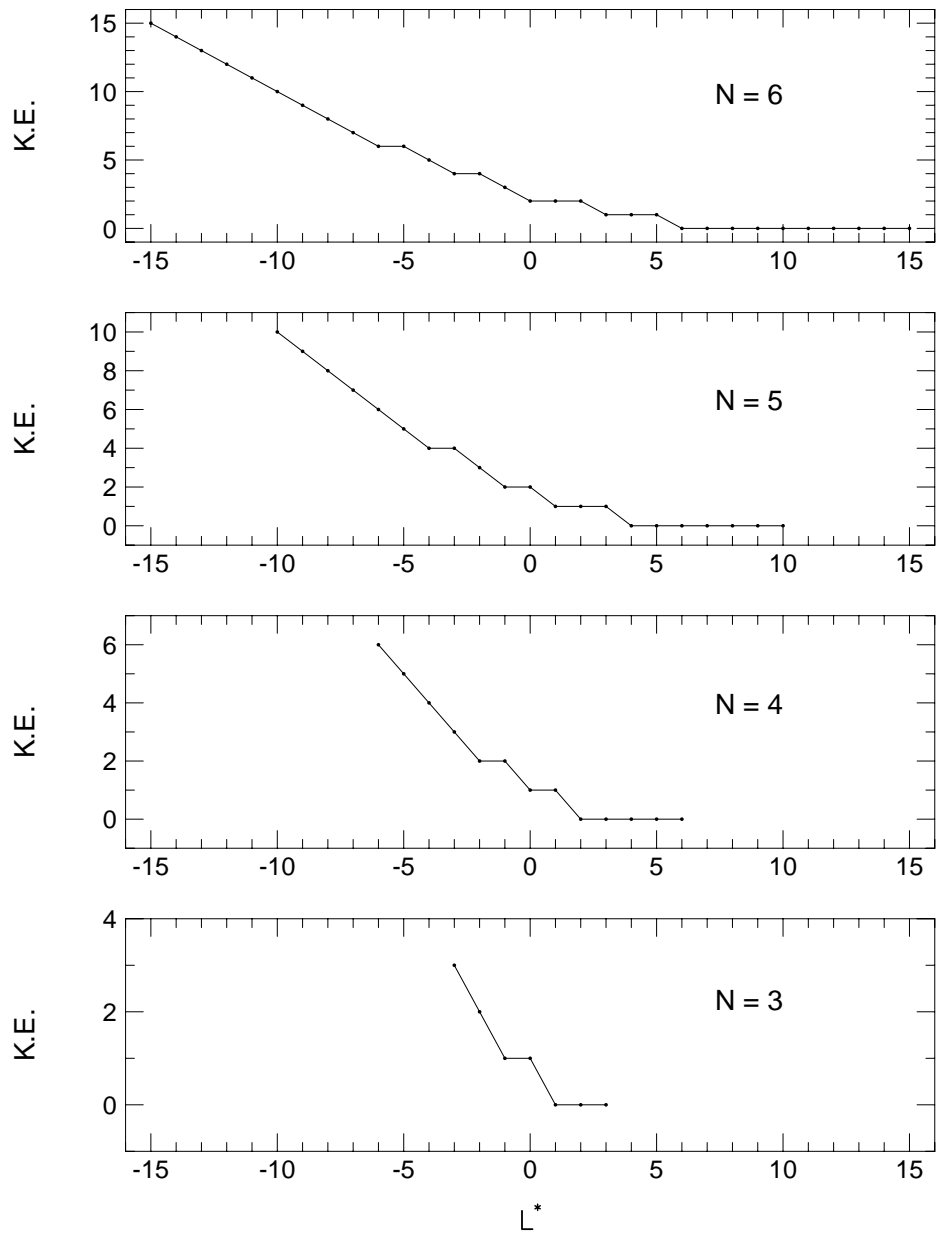
Fig. 6. The kinetic energy (K.E.) spectrum of non-interacting electrons
($N$=3,4,5,6)
with spin, in the range $-M \leq L^{*} \leq M$. The energies are given in
units of ${\hbar \omega^{CF}}$.

Fig. 7. (a) The low energy spectrum of the five electron system with Coulomb
interaction. (b)The low energy spectrum of a five fermion system with Coulomb
interaction. The energies are in units of ${e^2 \over \varepsilon a}$.


\begin{thebibliography}{99}
\bibitem{footnote} We remind the reader that in the $n$th LL
($n=0,1,...$) the single particle angular momentum can take
values $\ell=-n,-n+1,...$
\end{thebibliography}
\end{document}